\documentclass[bibyear]{aa} 

\usepackage{graphicx}
\usepackage{txfonts}
\begin{document}
   \title{Milky Way rotation curve from proper motions of red clump giants}

   \subtitle{}

   \author{
         Mart\'\i n L\'opez-Corredoira \inst{1,2}
          }
\institute{
$^1$ Instituto de Astrof\'\i sica de Canarias,
E-38200 La Laguna, Tenerife, Spain\\
$^2$ Departamento de Astrof\'\i sica, Universidad de La Laguna,
E-38206 La Laguna, Tenerife, Spain}

\offprints{martinlc@iac.es}
\titlerunning{Rotation curve / Red Clumps}
\authorrunning{L\'opez-Corredoira}

   \date{Received xxxx; accepted xxxx}

 
  \abstract
    {}  
  {We derive the stellar rotation curve of the Galaxy in the range of Galactocentric radii of $R=4-16$ kpc at different vertical heights from the Galactic plane of $z$ between -2 and +2 kpc. With this we reach high Galactocentric distances in which the kinematics is poorly known due mainly to uncertainties in the
distances to the sources.}
  {We used the PPMXL survey, which contains the USNO-B1 proper motions catalog  
cross--correlated with the astrometry and near-infrared photometry of the 2MASS Point Source
Catalog. To improve the accuracy of the proper motions,
we calculated the average proper motions of quasars to know their systematic shift from zero in this PPMXL survey, and we applied the corresponding correction to the proper motions of the whole survey, 
which reduces the systematic error.
We selected from the color-magnitude diagram $K$ vs. $(J-K)$ the standard candles corresponding
to red clump giants and used the information of their proper motions to build
a map of the rotation speed of our Galaxy.}
  {We obtain an almost flat rotation curve with a slight
decrease for higher values of $R$ or $|z|$. The most puzzling result is obtained for the farthest removed
and most off-plane regions, that is, at $R\approx 16$ kpc and $|z|\approx 2$ kpc, where a significant deviation from a null average proper motion ($\sim 4$ mas/yr) in the Galactic longitude direction for the
anticenter regions can be directly translated into a rotation speed much lower than at the solar
Galactocentric radius. In particular, we obtain an average speed of $82\pm 5$(stat.)$\pm 58$(syst.) km/s (assuming a solar Galactocentric distance of 8 kpc, and a circular/azimuthal velocity of 250 km/s for the Sun and of 238 km/s for the Local Standard of Rest), where the high systematic error bar is due mainly to the highest possible contamination of non-red clump giants and the proper motion systematic uncertainty.}
  {A scenario with a rotation speed lower than 150 km/s in these farthest removed
and most off-plane regions of our explored zone is intriguing, and invites one to reconsider different possibilities for the dark matter distribution. However, given the high systematic errors,
we cannot conclude about this. Hence, more measurements of the proper motions at high $R$ and $|z|$ are necessary to validate the exotic scenario that would arise if this low speed were confirmed.}


   \maketitle
%

\section{Introduction}

The rotation curve of the Milky Way disk was obtained using many different tracers and within
different Galactocentric distance ranges. 
Dias \& L\'epine (2005) determined the rotation speed of the spiral pattern of the Galaxy by direct observation of the birthplaces of open clusters of stars in the Galactic disk as a function of their age, confirming that the spiral arms rotate like a rigid body, as predicted by the classical theory of spiral waves. Bobylev et al. (2008) used space velocities of young open star clusters and the radial velocities of HI clouds and star-forming regions to derive the Galactic rotation curve in the range of $3<R($kpc$)<12$;
a more recent study by Reid et al. (2014) used proper motions of in-plane
star-forming regions in radio and reached $R\approx 16$ kpc.
Bovy et al. (2012) measured a flat rotation curve in the range $4<R($kpc$)<14$ by fitting the 
radial velocities of some stars. Compilations of other rotation curve measurements are given by Sofue et al. (2009) and Bhattacharjee et al. (2013) and references therein.

In general, the outer rotation curve of the stellar population is more poorly determined mainly because of the inaccuracy in the distance determination to the stars or other sources (Sofue 2011), 
and we wish to improve this
here. We use the red clump giant (RCG) population, which is a very appropriate standard candle (e.g., Castellani et al. 1992) with relatively low error bars on its distance estimate, and can be observed up to distances of $\approx 8$ kpc, which allows one to reach $R\lesssim 16$ kpc (L\'opez-Corredoira et al. 2002). Williams et al. (2013) have used RCGs to derive this kinematical information, but only within the range  $6<R($kpc$)<10$, $|z|<2$ kpc, which does not reach the farthest regions of the stellar disk we aim to explore here. To derive the Milky Way rotation curve from proper motions of RCGs in the range $4<R($kpc$)<16$ for $|z|<2$ kpc. 

The tool we used for our analysis of kinematics is a catalog of proper motions.
Bond et al. (2010) have also made an extended analysis of the stellar kinematics, including the rotation speed, from proper motions with a revised version of the catalog SDSS-DR7.
Aihara et al. (2011) reported later that there were significant problems with the measurement of the
astrometry of SDSS-DR8 and previous versions, which in turn leads 
to problems with their proper motions as well,
but Bond et al. (2010) may have taken them into account by analyzing the proper motion of quasars.
Nonetheless, this analysis of the SDSS-DR7 survey does not cover the Galactic plane at $|b|<20^\circ $, 
therefore it cannot be used to explore high values of $R$. 
Reid et al. (2014), as said, also analyze proper motions, but only for in-plane regions.
Here we concentrate on the region $|b|<20^\circ $.

The structure of this paper is as follows: in \S \ref{.data} the data source is described;
the method to subtract the RCG stars is detailed in \S \ref{.redclump}; 
the rotation curve from proper motions is calculated in \S \ref{.proper}, with an
extra section on how to correct for the systematic errors of proper motions in \S \ref{.syst}; this
leads to the results in \S \ref{.rotation} and their dynamical consequences in \S \ref{.dyn}.
A summary and conclusions are presented in the last section.
Unless specified otherwise we assumed for the Sun a Galactocentric distance of 8 kpc and a circular/azimuthal velocity of 250 km/s, and a circular/azimuthal velocity of the Local Standard of Rest (LSR) of 238 km/s.

\section{Data from the star catalog PPMXL: subsample of 2MASS}
\label{.data}

The star catalog PPMXL (Roeser et al. 2010) lists positions and proper motions of
about 900 million objects, aiming to be complete for the whole sky down to
magnitude $V\approx 20$. It is the result of the re-reduction of the catalog
of astrometry, visible photometry and proper motions of the USNO-B1 catalog  
cross--correlated with the astrometry and near infrared photometry of the 2MASS Point-Source
Catalog, and re-calculating the proper motions in the absolute reference frame of
International Celestial Reference Frame (ICRS), with respect to the barycenter
of the solar system. The typical statistical errors of these proper motion
is 4-10 mas/yr while the systematic errors are on average 1-2 mas/yr.
From the whole PPMXL, we first selected the subsample of 2MASS sources with 
$m_K\le 14$ and available $J$ photometry. This yielded 
a total of 126\,636\,484 objects with proper motions, i.e. 
an average of 3\,100 sources deg$^{-2}$.

\section{Selection of red clump giants}
\label{.redclump}

From these data, we selected the red clump giants (RCGs) in $K$ vs. $J-K$ color-magnitude
diagrams. Near-infrared color magnitude diagrams are very suitable for separating the
RCGs because of their clear separation of the dwarf population up to magnitude $m_K\lesssim 13$
(L\'opez-Corredoira et al. 2002). Moreover, the absolute magnitudes vary only little in the
near-infrared with metallicity or age (Castellani et al. 1992, Pietrzy\'nski et al. 2003). 
This narrow luminosity function distribution (Castellani et al. 1992) makes them very appropriate standard candles that trace the old stellar population of the Galactic structure.

\begin{figure}
\centering
\vspace{1cm}
\includegraphics[width=8cm]{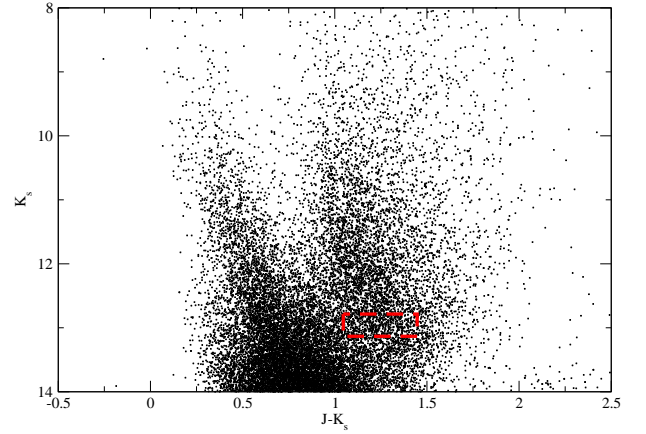}
\caption{Example of application of RCG selection from PPMXL survey in a $K$ vs. $J-K$ color-magnitude diagram.
This case is for the direction $\ell=125^\circ $, $b=2^\circ $, at one particular K-magnitude 
$m_{K,0}=13.0$, giving a maximum of counts
within RCG region at $(J-K)_0=1.25$. The dashed rectangle represents the area where the RCGs are selected.}
\vspace{1cm}
\label{Fig:CMl125b2}
\end{figure}

Here we used the same selection method for RCGs as in L\'opez-Corredoira et al. (2002).
For each magnitude $m_{K,0}$ we determined the $(J-K)_0$ which gives the highest count density along this horizontal line of the color-magnitude diagram for the RCG, and we considered all the stars within the rectangle with points $[(J-K), m_K]$ such that 
\begin{equation}
m_{K,0}-\Delta m_K \le m_K \le m_{K,0}+\Delta m_K
,\end{equation}\begin{equation}
Max[(J-K)_{{\rm red\ clump}}(J-K)_0-\Delta (J-K)]\le (J-K)
\end{equation}\[
\,\,\,\,\,\le (J-K)_0+\Delta (J-K)
.\] 
We take $\Delta m_K=\Delta (J-K)=$0.2 mag. Examples are given in Fig. \ref{Fig:CMl125b2} and in the top panel of \ref{Fig:l180b7}. 
Here, we ran the same application for the whole area of the sky with $|b|\le 20^\circ $ in bins of $\Delta \ell =1^\circ$, $\Delta b=1^\circ $ and $9.8<m_{K,0}\le 13.0$ in bins with $\Delta m_{K,0}$=0.4 mag. We did not investigate fields with $b>20^\circ $ because we are interested in the disk stars at low $z$; we did not not analyze either the brighter stars with $m_{K,0}<9.8$, because there are very few to accumulate statistics, nor the fainter stars with $m_{K,0}>13.0$ because they have higher ratio of dwarf contamination.
We selected only bins with more than ten sources. In total the number of RCGs with these criteria is 19\,189\,177 in an area of 14\,100 square degrees, that is, an average of 1\,360 RCGs per square degree.

Then, for each bin of RCGs with a given ($\ell $, $b$, $m_{K,0}$, $(J-K)_0$) we can
derive its cumulative extinction along the line of sight and its 3D position in the Galaxy 
($x$, $y$, $z$). The extinction and distance follow (L\'opez-Corredoira et al. 2002, \S 3):
\begin{equation}
A_K=0.67[(J-K)_0-(J-K)_{{\rm red\ clump}}]\ \ {\rm (Marschall\,et\,al.\,2006)}
,\end{equation}
\begin{equation}
r=10^{\frac{m_{K,0}-M_{K,{\rm red\ clump}}+5-A_K}{5}}
.\end{equation}
We used the updated parameter values: $M_{K,{\rm red\ clump}}=-1.60$, 
$(J-K)_{{\rm red\ clump}}=0.62$ (averages of the values given by Laney et al. 2012 and
Yaz G\"ok\c ce et al. 2013) and Galactocentric distance of the Sun $R_\odot =8$ kpc (Malkin 2013); 
we neglected the height of the Sun over the plane.
We did not consider any error in the distance determination because of the error in the absolute magnitude
of the RCGs, which is expected to be negligible in the outer disk: in the worst of the cases with $m_K=13.0$,
typical uncertainties of $\lesssim 0.02$ mag in the RCG absolute magnitude 
(Laney et al. 2012; Yaz G\"ok\c ce et al. 2013), plus the determination inaccuracy of the average
absolute magnitude on the order of 0.007 (for a typical number of stars per bin of 
400, and $\Delta m=0.2$), it leads
to relative errors in the distance of 1\% for distances of around 8 kpc, and 
gives a negligible error of $\sim 5$ km/s in the determination of the rotation curve (Sofue 2011, Fig. 5 multiplied by one half). For lower magnitudes the number of stars per bin is lower, which produces higher relative errors of the distance, but the error produced in the rotation curve is much lower (Sofue 2011, Fig. 5). We do not know the extinction uncertainty; assuming errors of up to 0.02 mag in K in the outer disk, we derive the same negligible uncertainties, except perhaps in the very few regions strictly in the plane ($z=0$) where the error might be somewhat larger.
In the inner disk, the errors may be larger, both because of higher extinction and because of
higher values of uncertainty of the rotation speed for a given relative error of distances (Sofue 2011, Fig. 5), but these are smaller than other statistical and systematic errors we derive.

\subsection{Contamination}
\label{.contam}

Our selected sources (almost 20 million) are mostly RCGs but there is also a small
fraction of contamination. These sources are:

\begin{itemize}

\item Extragalactic sources: they are on average $\approx 13$ galaxies/deg$^2$
up to magnitude $m_K=13.2$ (14.4 with $m_K\le 13.5$, L\'opez-Corredoira \& Betancort-Rijo 2004; with the
correction for a differential galaxy count of about five galaxies/deg$^2$/mag at $m_K=13.3$, 
Cole et al. 2002), with fluctuations $\frac{\delta N}{N}\approx 0.5$ in areas of 
one square degree (L\'opez-Corredoira \& Betancort-Rijo 2004). This is negligible
in comparison with the density of stars. Nonetheless, the contamination is even much
lower because within this range of magnitudes most of them are extended objects, which
are not included in Point Source Catalog of 2MASS.
The most important extragalactic contamination stems from quasars, which are
$\lesssim 0.05$ deg$^{-2}$ up to magnitude $m_K=13.2$ (from the data by Wu et al. 2011; see \S 
\ref{.syst}), and there are fewer of them with the same colors as giant stars. This is again negligible compared with the average 1,360 RCG stars per square degree in our fields.
Therefore, we consider that this very small contamination does not affect our results
significantly.

\item Main-sequence stars: there is some contamination of
the dwarfs from the main sequence: for the faintest bin
of magnitude ($m_K=13$), $\lesssim 10$\% of the stars in our selected sample of RCGs might
indeed be main-sequence dwarfs (L\'opez-Corredoira et al. 2002, \S 3.3.3); for brighter magnitudes, the contamination is lower. Indeed, the contamination might be slightly lower because we added an extra-constraint: here $(J-K)>0.62$, which avoids the contact
with the main sequence in regions of low extinction, whereas in L\'opez-Corredoira et al. (2002) the minimum of $(J-K)$ was 0.55. 

\item Other types of giant stars:
there may be some giants that are not RCGs in our
selected sample: M giants, asymptotic giant branch bump stars and 
red giant branch bump (RGBB) stars (Nataf et al. 2013). They are negligible except for
the third one (RGBB), which has absolute magnitudes $M_K$ between +2.0 and -1.4 
(L\'opez-Corredoira et al. 2002, \S 3.3.2) with a peak around 0.8 mag fainter than the RCG peak 
(Wegg \& Gerhard 2013; they analyzed the RCG for the bulge, but there are no significant differences
with the RCG luminosity function in the disk). According to Wegg \& Gerhard (2013),
the luminosity function at the observed color yields about 20\% RGGB stars with respect to the 
RCG stars at a given distance (this is a 16.7\% of the total of RGGB+RCG). 
Because the distances of both populations are related by $r_{\rm RGGB}=0.7r_{RCG}$
(derived from the difference of 0.8 mag) and the number of observed stars at a given distance is 
proportional to $N(r)=N(1)r^3\rho (r)$, where $N(r)$ is a monotonously decreasing function
(because the number of RCGs increases with magnitude; see
Figs. \ref{Fig:CMl125b2} or in the top panel of \ref{Fig:l180b7}. Usually, $\frac{N(0.7r)}{N(r)}$ is between 0.4 and 0.7) and we derive a contamination of RGBBs of $\lesssim 10$\% as well.

\end{itemize}

The contamination of the last two items is
important for calculating the average proper motions, because dwarfs are much closer
than RCGs, and attributing an incorrect distance by a factor $f$ would imply multiplying its
real proper motion by a factor $f$, which in turn would mean 
huge linear velocities for these dwarfs and a considerably increased
average if there is some net average velocity of these dwarfs. 
But this can be improved by using the median of the proper motions instead of
the average of proper motions and by bearing in mind that there might be 
a systematic error due
to this contamination, which would move the median to the position of the ordered set of proper motions
in the range of 40-60\% instead of 50\%. See the discussion in \S \ref{.dyn}. 

\subsection{Proper motion of the RCGs}

From the mean 3D position of the bin with ($\ell $, $b$, $m_{K,0}$) with a number $N$ of RCG stars, we calculated its median proper motion in Galactic coordinates: $\mu _\ell $, $\mu _b$. 
The angular proper motion is, of course, directly converted into a linear velocity
proper motion ($v_\ell $, $v_b$) simply by multiplying $\mu _\ell \cos b$ and $\mu _b$ by its distance from the Sun. Only bins with $N>10$ are included in the calculation. We found the use of the median instead of the average more appropriate because, as said in the previous subsection, it is a better way to exclude the outliers because of all kind of errors. 
The statistical error bars of the median were calculated for a 95\% C.L.: the upper and lower limits 
correspond to the positions $0.5N\pm 0.98\sqrt{N}$ of the ordered set of $N$ data. 
Since the errors of the averages are evaluated by a $\chi ^2$ analysis, the confidence level associated to these error bars is not important at this stage; their inverse
square was just used as weight in the weighted averages of multiple bins and the error bar of these
averages were quantified from the dispersion.
Moreover, as said above, we calculated a systematic error due to contamination: the upper and lower limits 
correspond to the positions $0.5N\pm 0.1N$ of the ordered set of $N$ data (assuming the worst cases
in which the contamination of non-RCGs is 20\% and that the proper motions of these non-RCGs are all higher
or lower than the median).

\section{Deriving the rotation curve from proper motions}
\label{.proper}

\begin{figure}
\centering
\includegraphics[width=8cm]{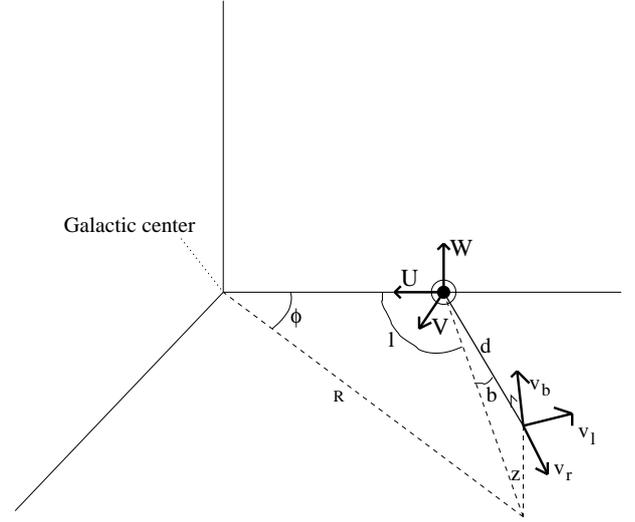}
\caption{Representation of kinematics of a star with respect to the Sun.}
\vspace{1cm}
\label{Fig:convvel}
\end{figure}

The 3D velocity of the combination of radial velocity ($v_r$) and proper motions ($v_\ell $, $v_b$)
is related to the velocity in the reference system $U$, $V$, $W$ as plotted in Fig. \ref{Fig:convvel}
by
\begin{equation}
v_r=U_*\cos \ell \cos b+V_*\sin \ell \cos b+W_*\sin b
,\end{equation}\[
v_\ell =-U_*\sin \ell+V_*\cos \ell
\]\[
v_b=-U_*\cos \ell \sin b-V_*\sin \ell \sin b+W_*\cos b
,\]
where $(U_*, V_*, W_*)$ is the velocity of a star relative to the Sun in the system (U,V,W).
The Sun velocity in this system with respect to the Galactic center
is $(U_\odot , V_{g,\odot }, W_\odot )$; the second coordinate
\begin{equation}
V_{g,\odot }=V_c(R_\odot ,z=0)+V_\odot 
,\end{equation}
where $V_c(R_\odot ,z=0)$ is the rotation speed of the Local 
Standard of Rest with respect to the Galactic center;
$(U_\odot , V_\odot , W_\odot )$ is the velocity of the Sun with respect to the LSR.
Here, we adopted the values $U_\odot =14.0\pm 1.5$ km/s, $V_\odot =12\pm 2$ 
km/s and $W_\odot =6\pm 2$ km/s (Sch\"onrich 2012). We used two values of $V_{g,\odot }$:
$250\pm 9$ km/s (Sch\"onrich 2012) and 200 km/s (the minimum value in the literature;
see McMillan \& Binney 2010, Bhattacharjee et al. 2013).

Assuming average circular orbits with rotation speed $V_c(R,z)$ at Galactocentric radius $R$, independent
of the azimuth $\phi $, the coplanar velocities of the star with respect to the Sun are
\begin{equation}
\label{UV}
U_*=-U_\odot +V_c(R,z)\sin \phi
,\end{equation}\[
V_*=-V_{g,\odot }+V_c(R,z)\cos \phi
,\]
where $\phi $ is the Galactocentric azimuth of the star. We did not consider an asymmetric drift
in each orbit (e.g., Golubov et al. 2013) because of the differences with perfect circular orbits: we cannot
determine with enough precision the intrinsic velocity dispersion necessary for calculating
the asymmetric drift, because
the measurement errors in the proper motions dominate, and, in any case, the expected correction value
is lower than 5\% (Bovy et al. 2012, Fig. 4 bottom panel), which is lower than our typical error bars for the
rotation speed. We did not consider the Galactic warp either: if we had considered
the warp modeled as a set of circular rings that are rotated and whose orbit is in a plane
with angle $i_w(R)$ with respect to the Galactic plane, then the terms $V_c(R,z)$ in Eq. (\ref{UV})
would need to be multiplied by a factor $\cos [i_w(R)\cos(\phi -\phi _w)]$, where $\phi _w$ is the azimuth
of line of nodes. The maximum elevation of the stellar warp at $R=16$ kpc is $z\sim 1$ kpc (Reyl\'e et al.
2009), which gives $i_w(R)=\tan ^{-1}\left(\frac{1\ {\rm kpc}}{16\ {\rm kpc}}\right)\sim 4^\circ $ and consequently $\cos [i_w(R)\cos(\phi -\phi _w)]>0.998$, which is well approximated by unity, that is, neglecting
the warp.

Hence,
\begin{equation}
V_c(R,z)=\frac{v_\ell -U_\odot \sin \ell +V_{g,\odot }\cos \ell}{\cos (\phi +\ell)}
\label{velrot}
.\end{equation}
This is a generalization for any velocity of the Sun of Eq. (13) in Sofue (2011) for a null
velocity of the Sun with respect to Local Standard of Rest [note that in Sofue 2011
$s=-R\cos (\phi +\ell )$]. 

\begin{figure}
\vspace{1cm}
\centering
\includegraphics[width=8cm]{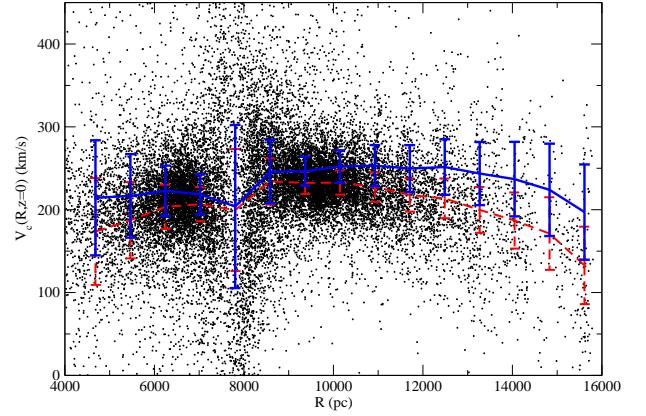}
\caption{Rotation speed as a function of $R$ for bins with $|z|<200$ pc, 
adopting a solar azimuthal speed of $V_{g,\odot }=250$ km/s and an LSR azimuthal speed
of $V_c(R_\odot ,z=0)=238$ km/s. The dashed line
is the weighted average of the bins without correction of systematic errors of the proper motion
but taking into account the errors due to contamination of non-RCGs
and including errors due to uncertainties in $(U_\odot , V_{g,\odot }, W_\odot )$.
The solid line is the weighted average of the bins including the correction of systematic errors 
of the proper motion as explained in \S \protect{\ref{.syst}}; error bars include statistical errors, uncertainties in $(U_\odot , V_{g,\odot }, W_\odot )$, errors due to contamination of non-RCGs
and errors of systematic errors of proper motion
(see Fig. \protect{\ref{Fig:errors}}).}
\vspace{1cm}
\label{Fig:rotcurve0}
\end{figure}

Eq. (\ref{velrot}) allows us to determine the rotation speed only with the determination
of the proper motion in the Galactic longitude projection. This is what we do here. 
For $z=0$, it is plotted in Fig. \ref{Fig:rotcurve0} for each bin, and a weighted average of
all the bins with common $R$, $z$.

\begin{figure}
\vspace{1cm}
\centering
\includegraphics[width=8cm]{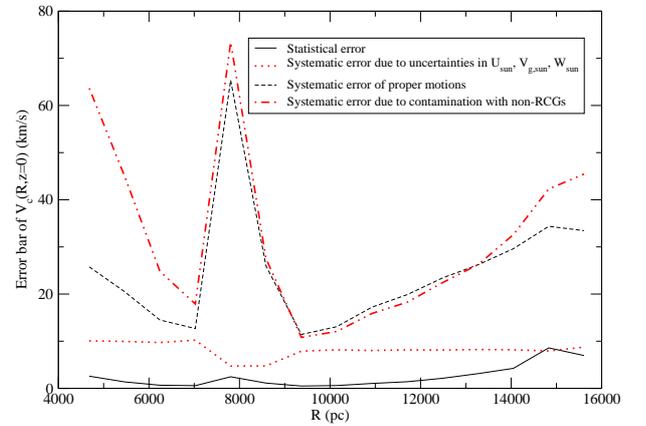}
\caption{Decomposition of the four sources of error, which sum quadratically to give
the error bars plotted in the solid line of Fig. \protect{\ref{Fig:rotcurve0}}.}
\vspace{1cm}
\label{Fig:errors}
\end{figure}

\section{Correcting systematic errors of the proper motions}
\label{.syst}

Proper motions published by the PPMXL catalog have both statistical and systematic errors.
The transmission of the statistical errors is taken into account in the previous steps; 
however, the systematic errors need to be accounted for as well because they are relatively high. In this subsection we calculate these systematic errors of the proper motions as a function of the Galactic coordinates.

Wu et al. (2011) analyzed the proper motions of 117,053 quasars within the PPMXL catalog. These
sources are supposed to have null proper motions, because of their very large distances, so their
average significant deviations from null proper motions can only stem from the systematic
errors in PPMXL. Other authors (e.g., Bond et al. 2010, \S 2.3.1; Dong et al. 2011) 
have also used quasars to evaluate systematic errors of proper
motions. Wu et al. (2011) found that the systematic errors depend on the coordinates, 
and they found no clear dependence with the magnitude or color of the sources. They
also found that the quasars that are included in the 2MASS survey (13\,520 sources) 
have a lower average systematic errors in the proper motion. 

We used here the same subsample from 2MASS as was used in subsection 4.4 of Wu et al. (2011): the 13,520 quasars. Because we also used sources that are both in PPMXL and the 2MASS survey, the systematic error in the proper motion of these quasars may well be applied to our sample of stars. 
Wu et al. (2011, \S 4.4) calculated an average $\mu _\alpha \cos (\delta )=-1.4$ mas/yr,
$\mu _\delta =-1.6$ mas/yr. Here we extend their analysis of the PPMXL-2MASS quasars, using Galactic
coordinates.

First, we confirm in this subsample of PPMXL-2MASS quasars the result given by Wu et al. (2011) for the whole set of PPMXL quasars: we found no significant correlation between proper motion
and magnitude. In particular for the K-magnitude, a weighted linear fit of the proper motions 
gives us a $\frac{d[\mu _\ell \cos (b)]}{dm_K}=+0.13\pm 0.10$ mas/yr/mag, $\frac{d\mu _b}{dm_K}=
-0.08\pm 0.10$ mas/yr/mag, which is compatible with no dependence or a negligible one.

\begin{figure}
\centering
\includegraphics[width=8cm]{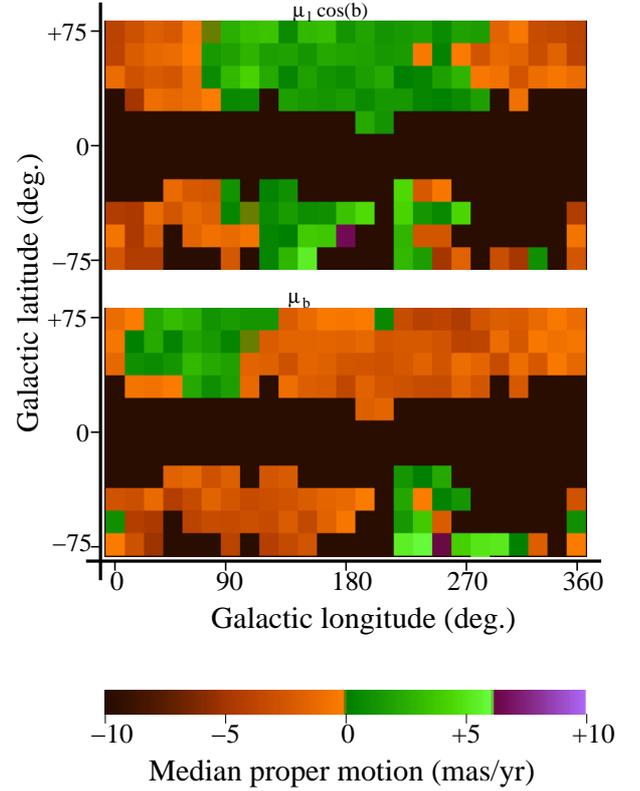}
\caption{Median of the proper motion of quasars included in the catalogs PPMXL and 
2MASS. Bin size: 15 deg.x15 deg. in Galactic coordinates.
Green to violet indicates positive values; brown to orange indicates negative
values; darkest colors indicate that there are fewer than 20 quasars in that bin, 
therefore no median was calculated.}
\vspace{1cm}
\label{Fig:QSOsystpos}
\end{figure}

There is, however, a clear dependence of the quasars proper motions on position. 
In Fig. \ref{Fig:QSOsystpos}, we plot the median of the proper motion 
in bins of $\Delta \ell=15$ deg., $\Delta b=15$ deg. in Galactic coordinates for bins 
with more than 20 quasars. 

Regrettably, there are not enough quasars in the Galactic plane, the zone in which we are interested.
Instead, we carried out a linear weighted interpolation in both directions of the map in Fig. \ref{Fig:QSOsystpos}, top panel, to derive the systematic error $Syst[\mu _\ell \cos b](\ell, b)$ with its corresponding statistical error. We took the error as the average error of the pixel in the surrounding area.
Then, we subtracted this systematic error from each of our proper motions of our bins:
\begin{equation}
(\mu _\ell \cos b)_{\rm corrected}=\mu _\ell \cos b - Syst[\mu _\ell \cos b](\ell, b)
\label{corrsyst}
,\end{equation}
and thus derived a rotation speed corrected for systematic errors of proper motions. See Fig.
\ref{Fig:rotcurve0} for the application in the $z=0$ cases. To calculate of the
error bars, the statistical errors are reduced in the weighted average in proportion to the root
square of the number of data but the errors of $Syst[\mu _\ell \cos b](\ell, b)$ are not, because
they are common for large regions.

One may wonder about the advantage of including the correction of Eq. (\ref{corrsyst}) if we
still keep a systematic error in the error of $Syst[\mu _\ell \cos b](\ell, b)$. The answer is that
the corrected values have an average null deviation, whereas the uncorrected proper motion does not. 
In Fig. \ref{Fig:rotcurve0} we illustrate how the correction shifts the rotation speed. See Fig.
\ref{Fig:errors} for the distribution of the error bar sources.

\section{Rotation curves}
\label{.rotation}

\begin{figure*}
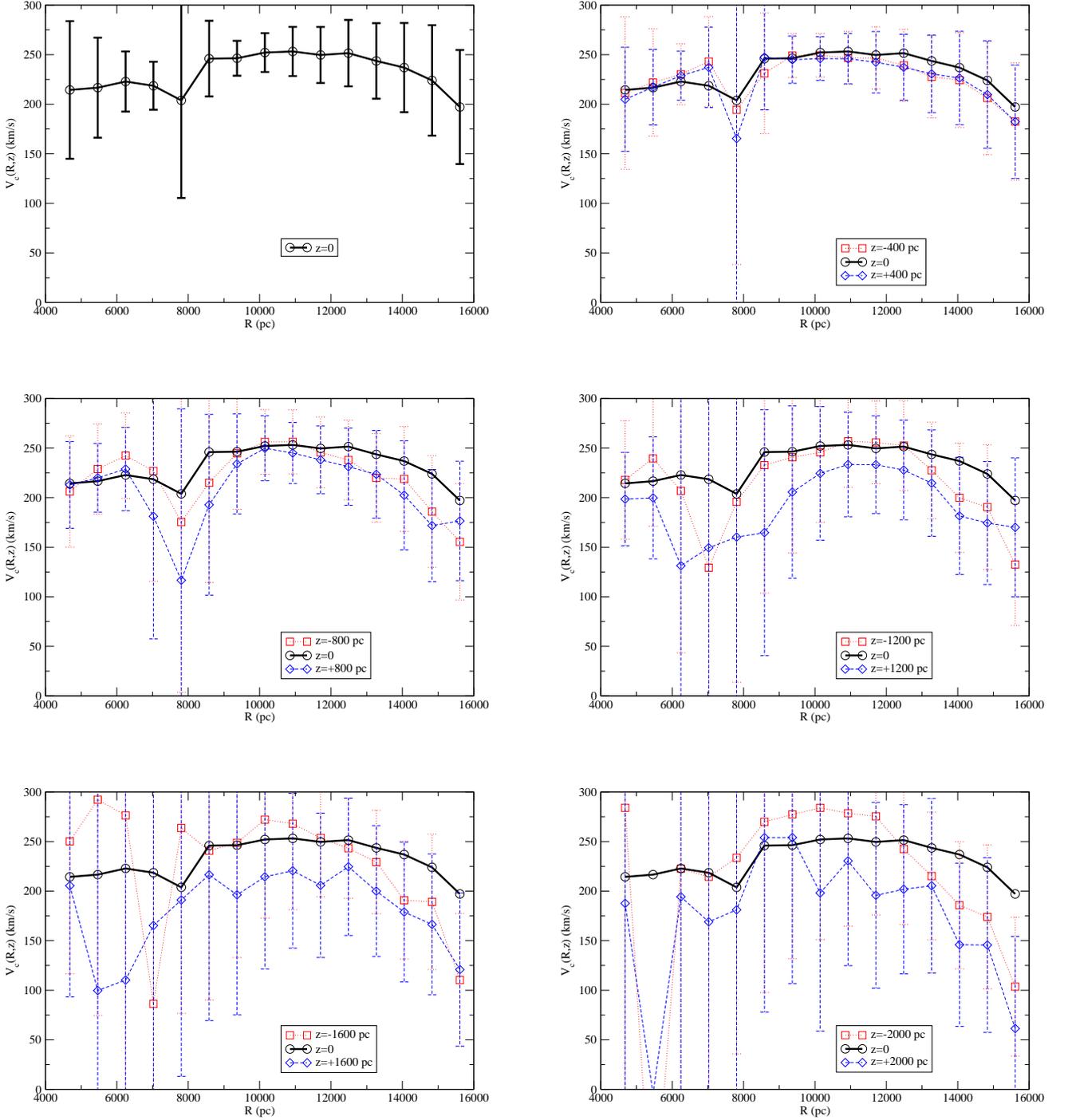

\centering
\includegraphics[width=8cm]{rotcurve0b.eps}
\hspace{1cm}
\includegraphics[width=8cm]{rotcurve1.eps}\\
\vspace{1cm}
\includegraphics[width=8cm]{rotcurve2.eps}
\hspace{1cm}
\includegraphics[width=8cm]{rotcurve3.eps}\\
\vspace{1cm}
\includegraphics[width=8cm]{rotcurve4.eps}
\hspace{1cm}
\includegraphics[width=8cm]{rotcurve5.eps}
\caption{Rotation speed derived from Eq. (\protect{\ref{velrot}}) as a function of $R$ for different $z$, with $\Delta z=400$ pc; correction of systematic errors of proper motions included; adopting a solar azimuthal speed of $V_{g,\odot }=250$ km/s and a LSR azimuthal speed of $V_c(R_\odot ,z=0)=238$ km/s. Error bars include statistical errors, uncertainties in $(U_\odot , V_{g,\odot }, W_\odot )$, and errors of the systematic errors of proper motion and contamination of non-RCGs (see Fig. \protect{\ref{Fig:errors}} for the distribution of errors for $z=0$).}
\vspace{1cm}
\label{Fig:rotcurve}
\end{figure*}

\begin{figure*}
\centering
\includegraphics[width=8cm]{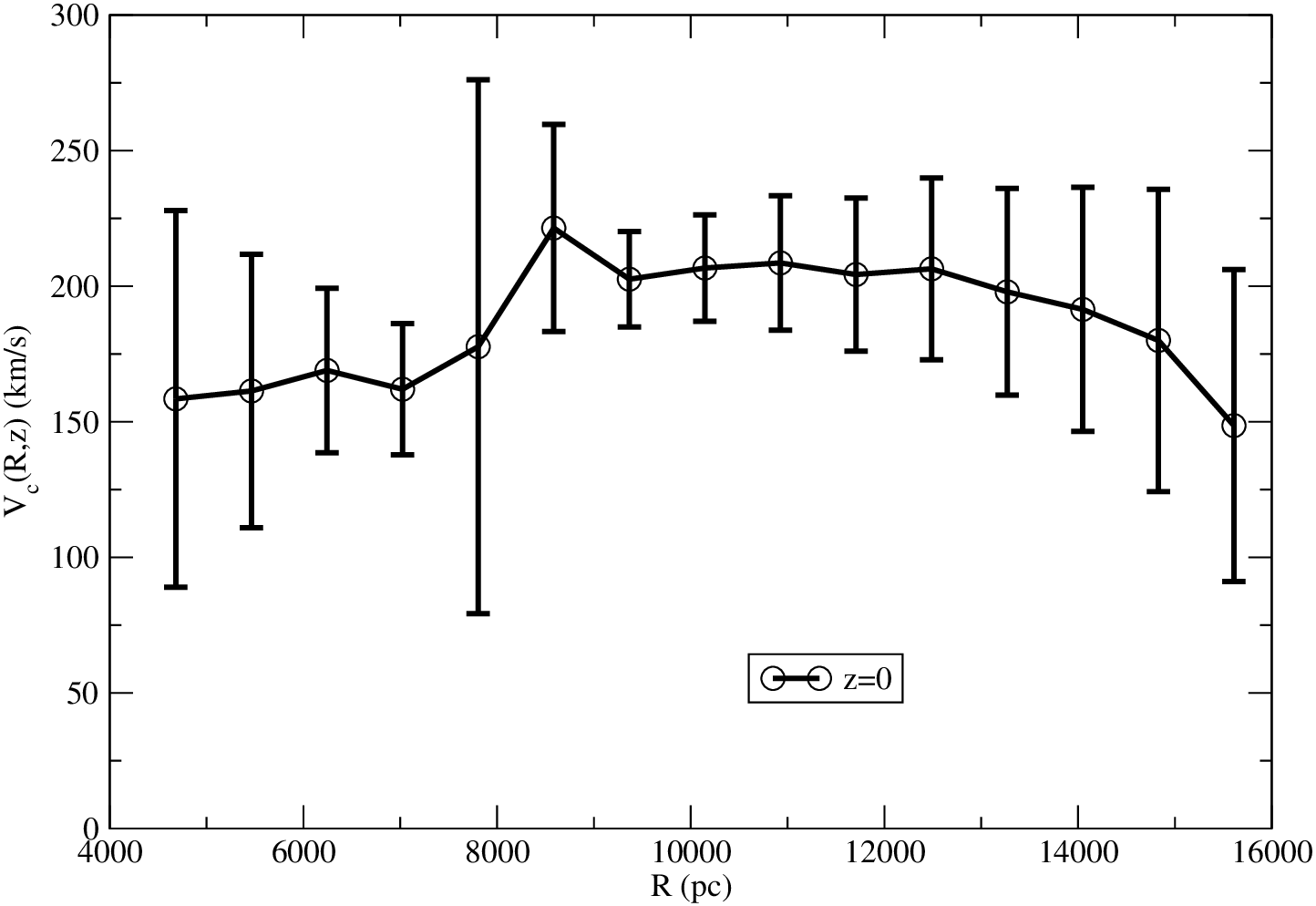}
\hspace{1cm}
\includegraphics[width=8cm]{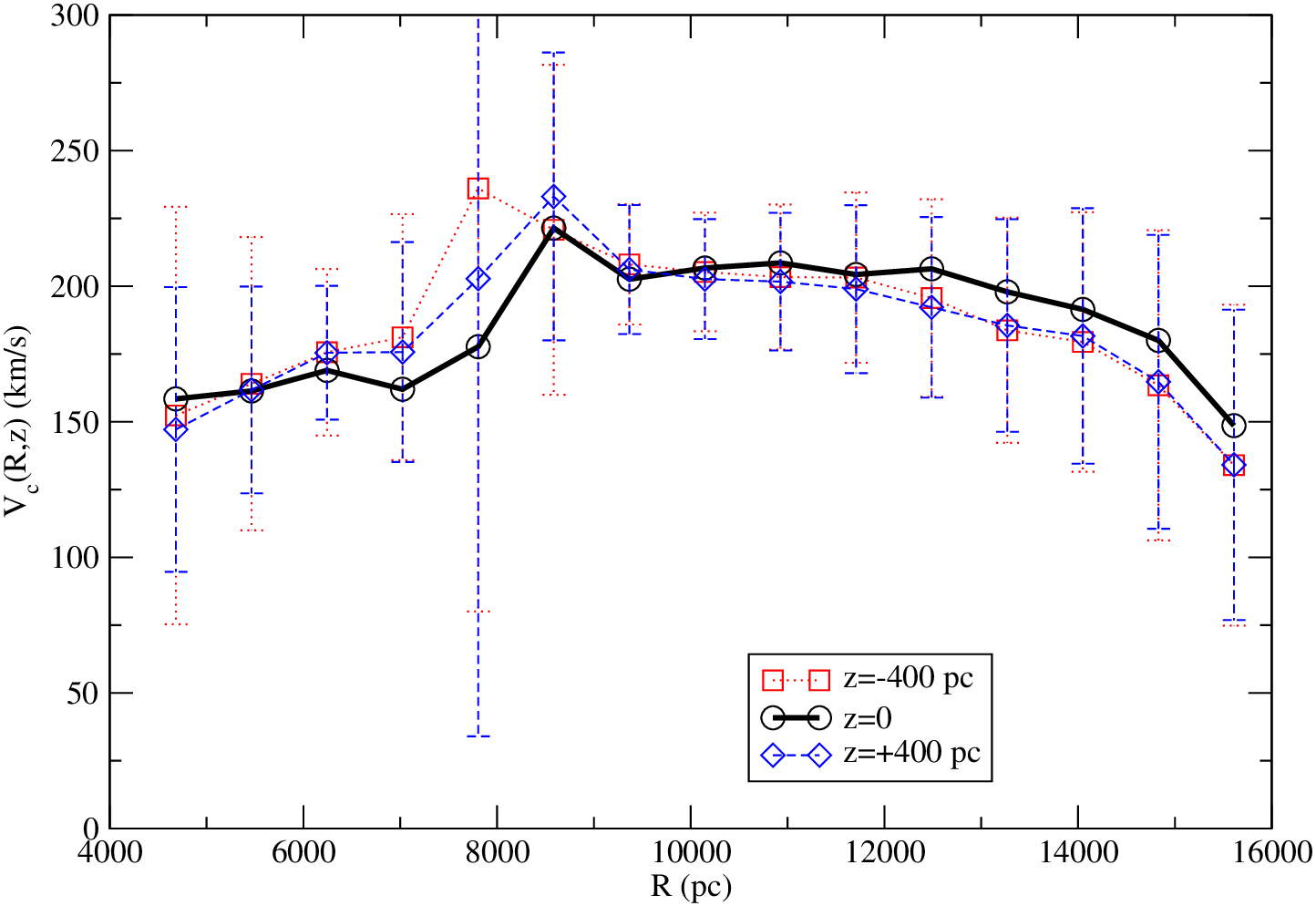}\\
\vspace{1cm}
\includegraphics[width=8cm]{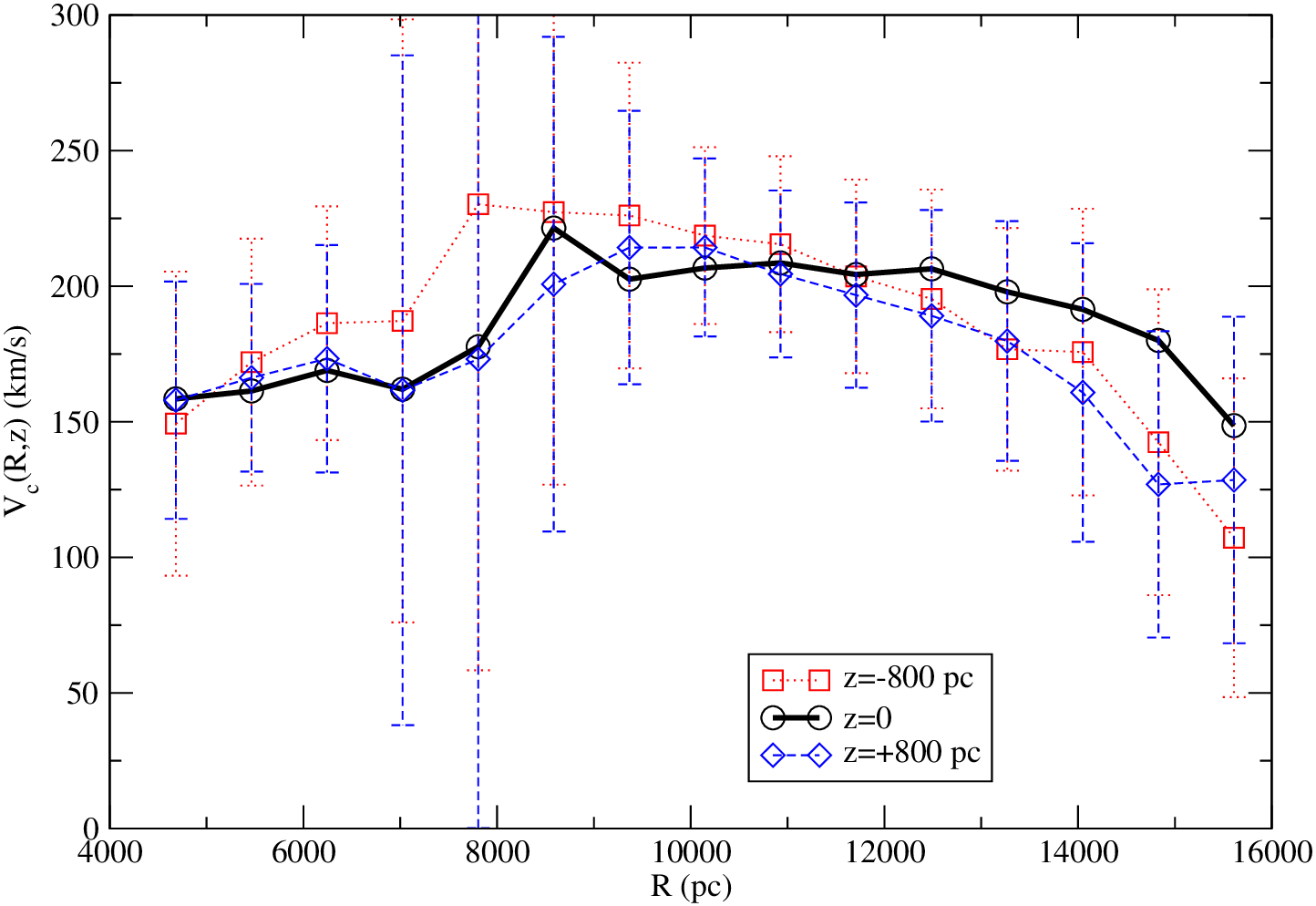}
\hspace{1cm}
\includegraphics[width=8cm]{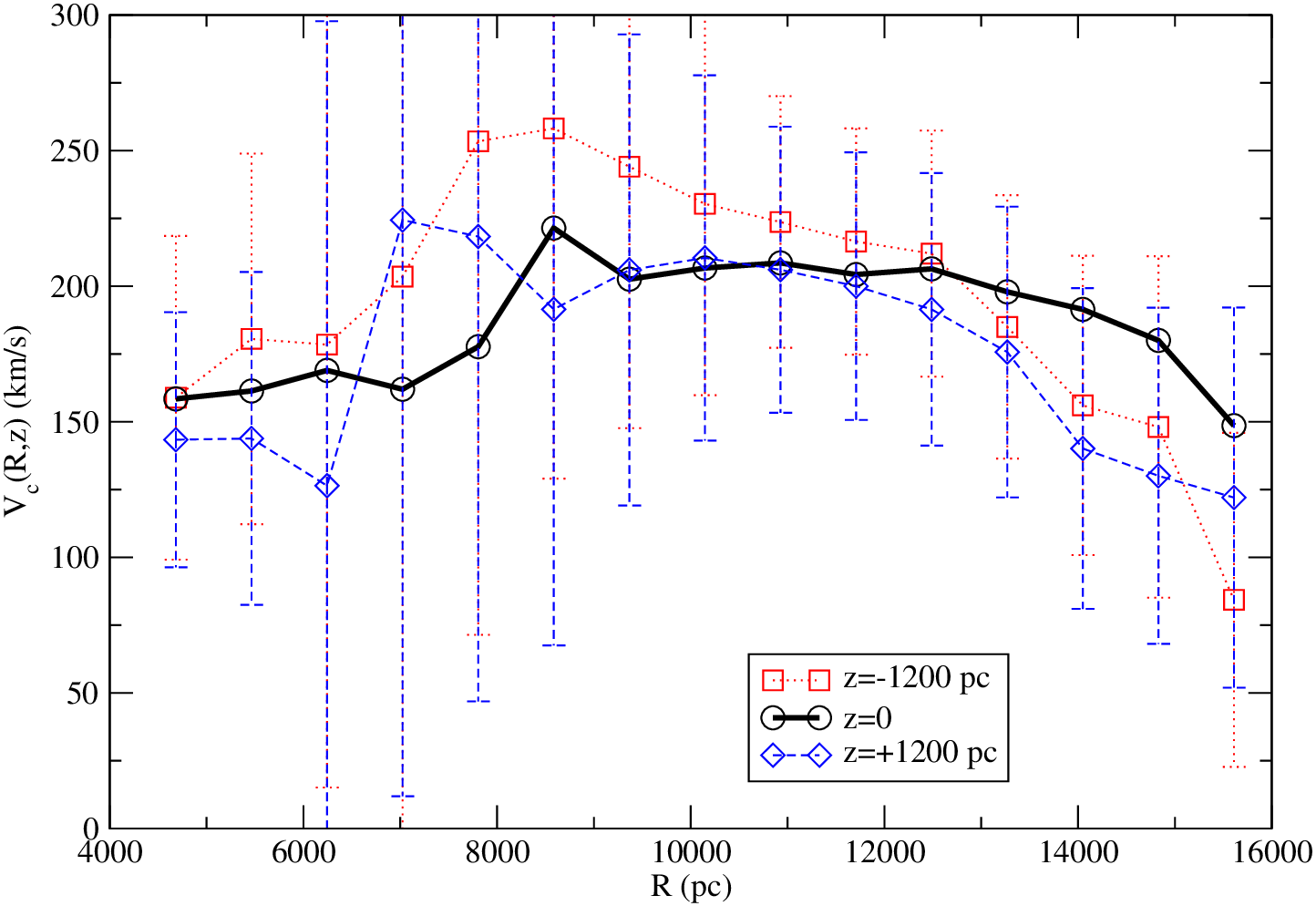}\\
\vspace{1cm}
\includegraphics[width=8cm]{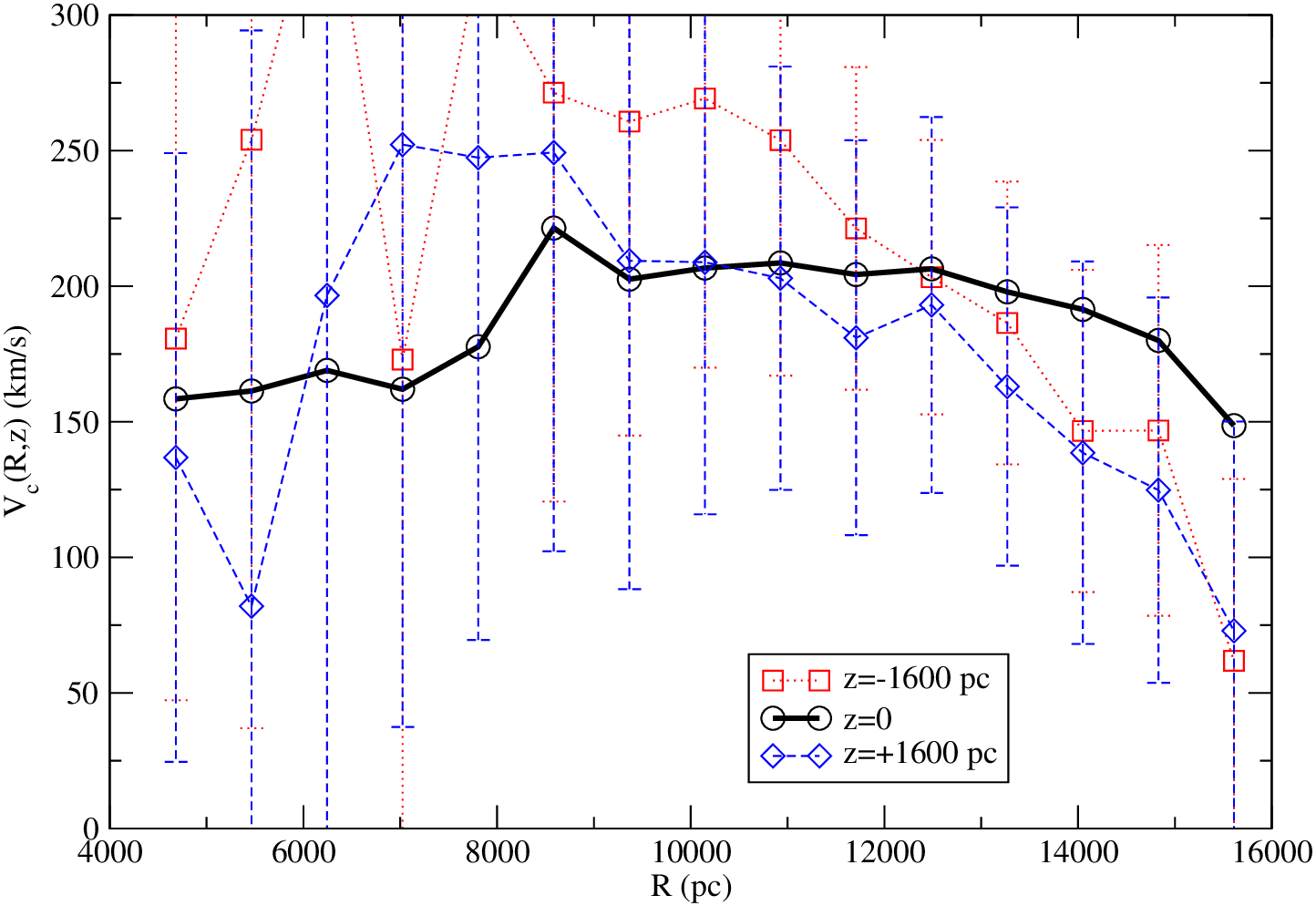}
\hspace{1cm}
\includegraphics[width=8cm]{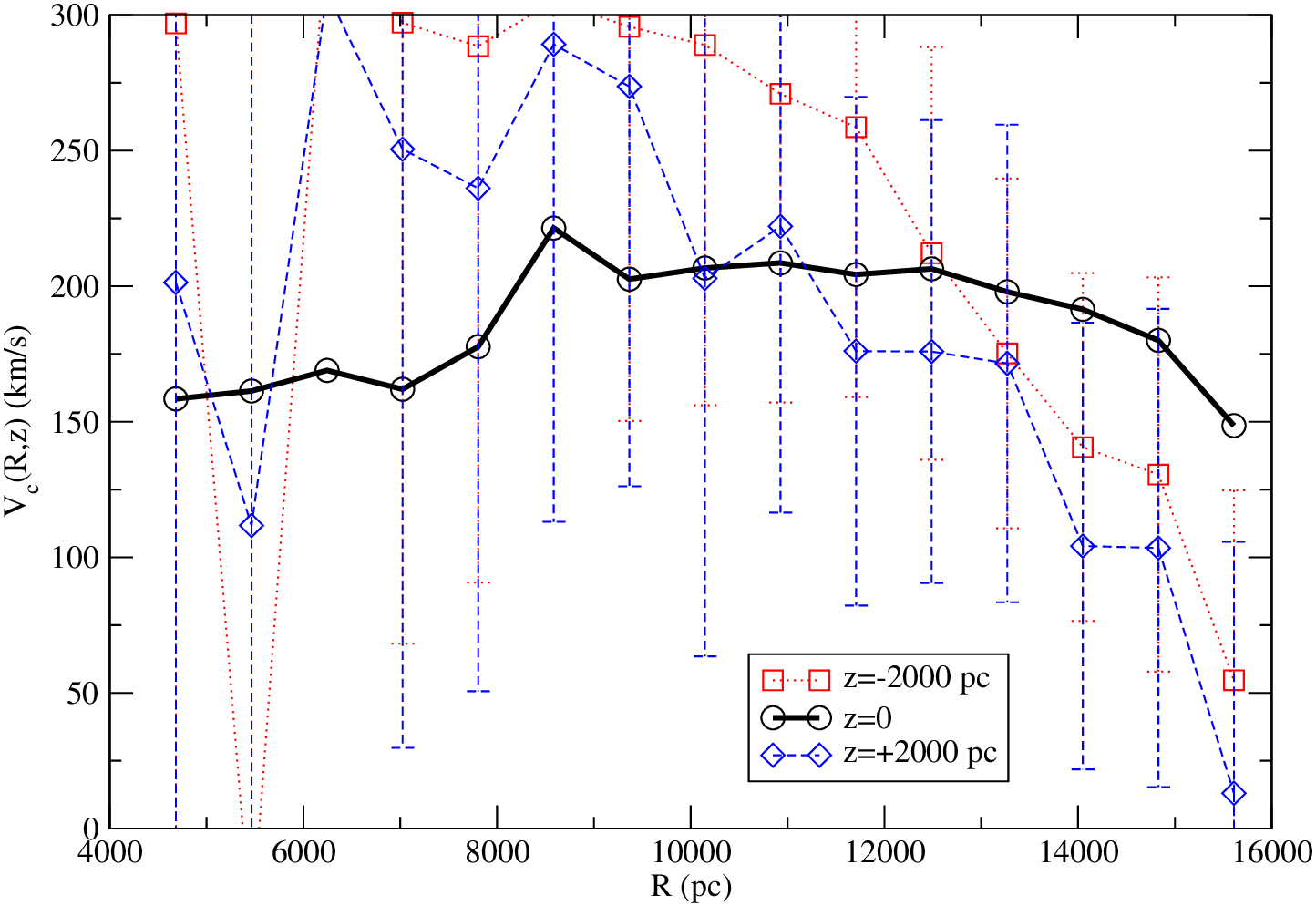}
\caption{Rotation speed derived from Eq. (\protect{\ref{velrot}}) as a function of $R$ for different $z$, with $\Delta z=400$ pc; correction of systematic errors of proper motions included; adopting a solar azimuthal speed of $V_{g,\odot }=200$ km/s and an LSR azimuthal speed of $V_c(R_\odot ,z=0)=188$ km/s. Error bars include statistical errors, uncertainties in $(U_\odot , V_{g,\odot }, W_\odot )$, and errors of the systematic errors of proper motion and contamination of non-RCGs.}
\vspace{1cm}
\label{Fig:rotcurve200}
\end{figure*}

We are interested in the region $R>4$ kpc, $|z|<2$ kpc, which defines the disk region. For $R<4$ kpc
we find the non-axisymmetric structure with non-circular orbits of the long bar (L\'opez-Corredoira et al. 2007). For $|z|>2$ kpc, the halo becomes important and the stellar density of the disk is a factor $\lesssim 1/300$ than at $z=0$ (Bilir et al. 2008). Fig. \ref{Fig:rotcurve} shows the rotation curves for different values of $z$  including the correction of systematic errors in the proper motions. Note that in these plots the error bars of the different bins are not entirely independent because the systematic errors are not independent. We used $V_{g,\odot }=250\pm 9$ km/s (Sch\"onrich 2012). If we had used the lowest value
in the recent literature of 200 km/s (McMillan \& Binney 2010, Bhattacharjee et al. 2013) [without any
error bar; at present, we are just interested to see the effect of the change of $V_{g,\odot }$ and
we do not think this is the correct value] and keeping $R_\odot =8$ kpc (although, this should be
lower if we reduce $V_{g,\odot }$; McMillan \& Binney 2010), we would have obtained similar results, Fig.
\ref{Fig:rotcurve200}, which shows us that the observed features in $V_c(R)$ are 
not very dependent on the value of $V_{g,\odot }$.
This is expected because, according to Eq. (\ref{velrot}), the rotation curves derived with two different values of $V_{g,\odot }$ are related by 
\begin{equation}
V_{c,2}(R,z)=V_{c,1}(R,z)+(V_{g,\odot ,2}-V_{g,\odot ,1})
\left\langle \frac{\cos \ell}{\cos (\phi +\ell )}\right\rangle (R,z)
.\end{equation}

The error bars are dominated by the systematic errors due to contamination of non-RCGs (we have assumed the most pessimistic scenario of a 20\% contamination and that the proper motions of these non-RCGs are all 
higher or lower than the median) and the systematic errors of the proper motions derived using the QSOs reference, see Fig. \ref{Fig:errors}. It can be appreciated that for the bins at $R\approx R_\odot $,
the error bar is much larger than others because
we are observing the velocity of the stars with a small
angle with respect to the line of sight and consequently with most of its contribution in the
radial direction and a small one to the proper motions. This implies that $\cos (\phi +\ell )$ in Eq. (\ref{velrot}) is small; we derive at $R\approx R_\odot $ values of $V_c$ 10-35 km/s lower than the assumed LSR circular speed, but they are perfectly consistent within the error bars.

Considering Fig. \ref{Fig:rotcurve}, with $V_{g,\odot }=250\pm 9$ km/s, the following features are observed in these rotation curves:
\begin{itemize}

\item In the Galactic plane ($z=0$), we observe an almost constant rotation curve for $R>R_\odot =8$ kpc,
although with a slight slope for $13<R($kpc$)<16$. Because of the large error bars mostly due to uncertainties
in the systematic effects, we cannot be sure that a flat rotation curve in this range is excluded.
A decreasing rotation speed for the largest $R$ of our range was also observed by other authors 
(Dias \& L\'epine 2005; Sofue et al. 2009). 
Nonetheless, for off-plane regions, this decreasing speed is more significant: 
for $z\gtrsim 1$ kpc we observe that highest rotation speed of $\approx 250$ km/s is reached at $R\approx 10$ kpc and then it decreases to values of $\approx 100$ km/s at $R\approx 16$ kpc, $z=2$ kpc.
 
\item In the inner Galaxy, with $R\lesssim R_\odot $, given the huge error bars, we observe no systematic variation of the rotation curve with $z$. Our results are consistent with lower rotation speeds at higher $z$ observed by other authors (Bond et al. 2010; Williams et al. 2013). 
For $13<R($kpc$)<16$, there is at $z\gtrsim 1$ kpc a lower rotation speed than at $z=0$. 
On average for the results with $V_{g,\odot }=250$ km/s and 
$R_\odot <R<2R_\odot $,
\begin{equation}
\frac{\partial V_c(R,z)}{\partial |z|}=2.0+2.4\,{\rm kpc}^{-1}\times (R-R_\odot)
\end{equation}\[\ \ \ \ \ \ 
-1.2\,{\rm kpc}^{-2}\times [(R-R_\odot )]^2\ {\rm km/s/kpc}
.\]

\item We observe some asymmetry between the northern and southern Galactic hemisphere, with the southern one with higher rotations speeds, more significantly in the range $8<R($kpc$)<13$. However, again, this fact is observed in some bins with the error bars.

\end{itemize}

\section{Dynamical consequences}
\label{.dyn}

The main challenge of our results is the low rotation velocity at the largest $R$, $|z|$ within our range.
Low rotation speed stars of $\approx 100$ km/s were previously 
observed for the mean rotation velocity of some fraction of stars attributed 
to the thick-disk population at $R\approx R_\odot $ for low z (Fuchs et al. 1999) or 
at $|z|\approx 2$ kpc (Gilmore et al. 2002). However, they were a minority of stars because
the average rotation curve we observe for these coordinates is 150-200 km/s, in agreement with
Bond et al. (2010, Fig. 14). If the average rotation curve decreases to low values
of $V_c$ this means, if the result were correct, that the mass distribution declines almost
in a Keplerian way ($\propto 1/R$).
The rotation curve beyond $2R_\odot $ was observed to possibly decline in Keplerian way
in the Galactic plane (Honma \& Sofue 1996), but not in off-plane regions like ours.

\begin{figure}
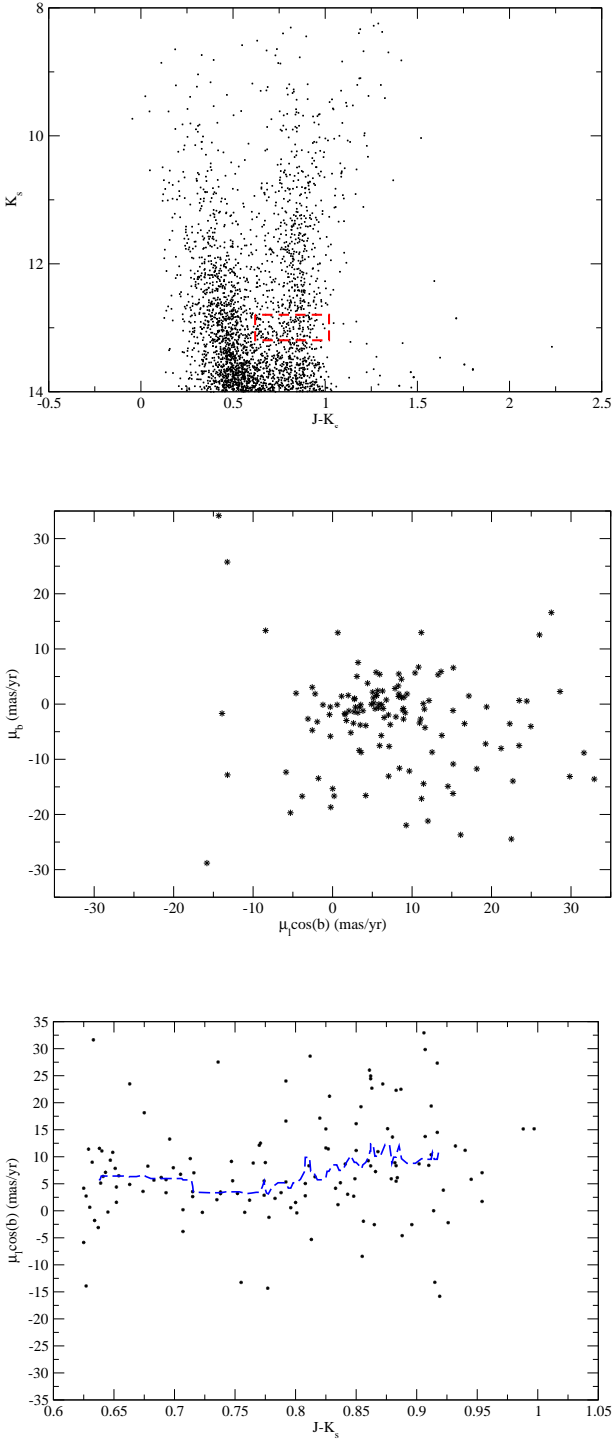

\centering
\vspace{1cm}
\includegraphics[width=8cm]{CMl180b7.eps}\\
\vspace{1cm}
\includegraphics[width=8cm]{pml180b7.eps}\\
\vspace{1cm}
\includegraphics[width=8cm]{pml180b7color.eps}\\
\caption{Upper panel: RCG selection in a $K$ vs. $J-K$ color-magnitude diagram
for the direction $\ell=180^\circ $, $b=7^\circ $, $m_{K,0}=13.0$, giving a maximum of counts
within the RCG region at $(J-K)_0=0.81$; the dashed rectangle represents the area where the RCGs are selected. Middle panel: proper motions of the selected RCG stars in the catalog PPMXL.
Bottom panel: proper motion in the direction of Galactic longitude vs. color; the 
dashed line indicates the median of 20 points.}
\vspace{1cm}
\label{Fig:l180b7}
\end{figure}

First we analyzed why our method leads to these low rotation speeds.
The regions with $R\approx 16$ kpc, $|z|<2$ kpc are around the anticenter ($\ell =180^\circ $)
and $|b|\lesssim 14^\circ $, with apparent magnitude of the RCG stars $m_K\approx 13.0$; there are
fewer bins than for other lower $R$ (see Fig. \ref{Fig:rotcurve0}).
If we take, for instance, the region $\ell =180^\circ $, $b=+7$ (with $\Delta \ell=1^\circ $,
$\Delta b=1^\circ $), we have 2,802 stars from the PPMXL/2MASS catalog with $m_K<14.0$; for
$m_K=13.0$ we determine a $(J-K)_0=0.81$, and the corresponding RCGs with
$12.8<m_K<13.2$, $0.62<(J-K)_0<1.01$ are 138 stars. Assuming that all these stars are RCGs, 
this corresponds to an average distance of 7.8 kpc
and an extinction of $A_K=0.13$, which is quite reasonable, and the selection of
RCGs looks appropriate. We can see in Fig. \ref{Fig:l180b7} the color-magnitude diagram
and the distribution of $\mu_\ell $ and $\mu _b$ of the selected RCGs. The median
of the $\mu_\ell \cos b$ is $6.8\pm 1.6$(95\% C.L.; statistical only)$\pm 1.8$ (maximum 
systematic error due to 20\% contamination) mas/yr.
Certainly, as discussed in \S \ref{.contam}, 
some of the points in Fig. \ref{Fig:l180b7}/middle might be contamination
of local dwarfs or RGBBs.
In Fig. \ref{Fig:l180b7} bottom panel, we see that these high proper motion cases, with either
positive or negative values, are almost randomly distributed in colors, showing only 
a slight trend: redder colors give slightly higher proper motions in the 
direction of the Galactic longitude. A selection of RCGs in a narrower range of colors
around $(J-K)_0=0.81$ would not give a result significantly different from our 
6.8 mas/yr. 
Now we correct for the systematic error in the proper motions: 
the bi-linear interpolation of the plot of Fig. 
\ref{Fig:QSOsystpos} is $Syst[\mu _\ell \cos b]=1.8\pm 0.9$ mas/yr and hence 
$(\mu_\ell \cos b)_{\rm corrected}=5.0\pm 1.8$(stat.)$\pm 1.8$ (syst.) mas/yr.
At the attributed distance, this corresponds
to a $v_\ell =180\pm 70$(stat.)$\pm 70$(syst.) km/s. 
In the anticenter ($\phi =0$, $\ell =180^\circ $), following
Eq. (\ref{velrot}), $v_\ell =V_c-V_{g,\odot }$, therefore a significant departure from zero leads
to a significant departure of the rotation speed with respect to the solar one. 
Are the systematic corrections of the proper motions are underestimated?
It is possible, because interpolations/extrapolations are always dangerous, but in
Fig. \ref{Fig:QSOsystpos} there is not too much variation of $Syst[\mu _\ell \cos b]$ in the
regions around our coordinate, $Syst[\mu _\ell \cos b]=1.8\pm 0.9$ mas/yr at $\ell =180^\circ $, $b=+7^\circ $
appears to be a reasonable interpolation. The five closest pixels in Fig. \ref{Fig:QSOsystpos} give
$2.1\pm 0.8$ mas/yr at $\ell =195^\circ $, $b=+15^\circ $; 
$1.1\pm 1.7$ mas/yr at $\ell =210^\circ $, $b=+15^\circ $;
$1.2\pm 1.1$ mas/yr at $\ell =165^\circ $, $b=+30^\circ $;
$1.5\pm 0.6$ mas/yr at $\ell =180^\circ $, $b=+30^\circ $;
$0.6\pm 0.4$ mas/yr at $\ell =195^\circ $, $b=+30^\circ $, which are all close to the
interpolated value of $1.8\pm 0.9$.
This is only for
one bin, the represented points in Figs. \ref{Fig:rotcurve} and \ref{Fig:rotcurve200} stand for a weighted
average of several bins like this.  To conclude,
the origin of this strong decrease in rotation speeds is clear:
the high values of $(\mu_\ell \cos b)_{\rm corrected}$ in off-plane regions of the anticenter. 
We obtain rotation speeds of $V_c\approx 100$ km/s because we derive average $(\mu_\ell \cos b)_{\rm corrected}\approx 4$ mas/yr; the only possibility to make this number lower is considering that the 
systematic errors of 2-3 mas/yr are in the direction of reducing this proper motion toward 1-2 mas/yr, 
which would mean a $V_c=175$ km/s, which would agree better with expectations.
At present, with the obtained results and the associated error bars, we cannot exclude this possibility.

A strong vertical  gradient in $V_c$ would mean that the disk contribution is the most important one in the rotation curves for $R\le 16$ kpc. Since the radial acceleration
\begin{equation}
a_R(R,z)=a_{R,disk}(R,z)+a_{R,bulge+bar}(R,z)+a_{R,halo}(R,z)=\frac{V_c^2(R,z)}{R}
\end{equation}
is very dependent on $z$ for $13<R$(kpc)$<16$, and $a_{R,bulge+bar}$ and $a_{R,halo}$ are almost
independent of $z$, we would deduce that the dominant component here is $a_{R,disk}$.
The component of mass distribution corresponding to the bulge+long bar is within $R<4$ kpc and, although they are not spheroids, at Galactocentric distances of $13<R$(kpc)$<16$ the acceleration can be 
approximated by the monopolar Keplerian term:
\begin{equation}
a_{R,bulge+bar}(R,z)\approx \frac{GM_{bulge+bar}R}{(R^2+z^2)^{3/2}}
,\end{equation}
for which 
\begin{equation}
\frac{1}{a_{R,bulge+bar}(R,z)}\frac{\partial a_{R,bulge+bar}(R,z)}{\partial z}\approx -\left(\frac{3z}{R^2}\right)\ll 1
,\end{equation} 
thus the vertical gradient is negligible. Moreover $M_{bulge+bar}\sim 2\times 10^{10}$ M$_\odot$ (Sevenster et al. 1999; L\'opez-Corredoira et al. 2007), which is only a low ratio
of the mass of the Galaxy. 
For the halo, the total mass ratio is very high but most of contribution comes from the outest Galaxy,
much farther away than $R=16$ kpc; a calculation of $a_{R,halo}$ using an oblate spheroid of major axis $a$
and vertical axis $c<a$ with a density distribution $\rho \propto \frac{1}{s^2(s^2+s_0^2)^{3/2}}$ 
(Battaglia et al. 2005) with $s_0=105$ kpc and $s=\sqrt{R^2+\frac{a^2}{c^2}z^2}$, we compute
for $0.6<\frac{c}{a}<1.0$ that 
\begin{equation}
\frac{1}{a_{R,bulge+bar}(R,z)}\frac{\partial a_{R,bulge+bar}(R,z)}{\partial z}\lesssim \frac{z}{(9.3\ {\rm kpc})^2}\ll 1
,\end{equation} 
a negligible relative gradient in $z$; other models of the halo also give results on the same order. 
Therefore, only a disk with a strong non-monopolar term could give a contribution with a strong
vertical gradient in $V_c$, and this would correspond to high concentrations of mass in the outer disk. 
It was proved long time ago (Kuzmin 1952, 1955) that there is no evidence for the presence of large amounts dark matter in the disk of the Galaxy from observations in the solar neighborhood, and this is still confirmed today (Garbari et al. 2012), 
but one might investigate whether amounts of dark matter in the outer
disk are possible. From the theoretical point of view at least, it is a general solution for the observed rotation curves: Feng \& Gallo (2011) showed that some mass distributions of the disk alone might mimic the flat rotation curves of a galaxy (Mestel's disk), without dark matter halos, although these distributions are not realistic compared with the observed stellar distribution (exponential).

\section{Summary and conclusions}
\label{.concl}

Using only proper motions of sources identified as RCGs, 
we have derived the stellar rotation curve of the Galaxy in the range of Galactocentric radii of $R=4-16$ kpc for different vertical shifts from the Galactic plane of $z$ between -2 and +2 kpc. We obtained an almost flat rotation curve with a slight
fall-off for higher values of $R$ or $|z|$. The most puzzling result is the observed low average rotation speed at the outest and most off-plane cases, that is, at $R\approx 16$ kpc and $|z|\approx 2$, with less than the half of the rotation speed with respect to its value in the solar neighborhood region: on average for $z=\pm 2$ kpc, $V_c=82\pm 5$(stat.)$\pm 58$(syst.) km/s (assuming the systematic errors at $z=-2$ kpc and $z=+2$ kpc are independent, so they sum quadratically). If this speed is lower than 150 km/s (for $V_{g,\odot }=250$ km/s), it would have important consequences for the analysis of the dark matter 
distribution in the outer part
of our Galaxy. Nevertheless, at present we cannot exclude that the strong vertical gradient in the rotation speed is caused only by a combined systematic error of proper motions and contamination with non-RCGs. 
The assumptions made here, such as the approximation of circular orbits, might also be relaxed in the outest part of the disk. More measurements of the proper motions at high $R$ and $z$ are warranted to confirm
or refute these results.

\begin{acknowledgements}

Thanks are given to F. Garz\'on and the anonymous referee for helpful comments and suggestions
to improve this paper. Thanks are given to Astrid Peter (language editor of A\&A)
for proof-reading of the text.
The author was supported by the grant AYA2012-33211 of the Spanish Science Ministry.
This work has used the data of PPMXL catalog (Roeser et al. 2010). Z.-Y. Wu has kindly provided the data corresponding to the results of the paper Wu et al. (2011). 
\end{acknowledgements}


\begin{thebibliography}{99}

\bibitem{} Aihara, H., Allende Prieto, C., An, D., et al. 2011, ApJS, 195, 26 

\bibitem{} Battaglia, G., Helmi, A., Morrison, H., et al. 2005, MNRAS, 330, 35

\bibitem{} Bhattacharjee, P., Chaudhury, S., \& Kundu, S. 2013, arXiv:1310.2659

\bibitem{} Bilir, S., Cabrera-Lavers, A., Karaali, S., Ak, S., Yaz, E., \& L\'opez-Corredoira, M.
2008, PASA, 25, 69

\bibitem{} Bobylev, V. V., Bajkova, A. T., \& Stepanishchev, A. S. 2008,
Astron. Lett., 34, 515

\bibitem{} Bond, N. A., Ivezi\'c, Z., Sesar, B., et al. 2010, ApJ, 716, 1

\bibitem{} Bovy, J., Allende Prieto, C., Beers, T. C., et al., 2012, ApJ, 759, 131

\bibitem{} Castellani V., Chieffi A., \& Straniero O., 1992, ApJS, 78, 517

\bibitem{} Cole, S., Norberg, P., Baugh, C., et al. 2001, MNRAS, 326, 255

\bibitem{} Dong, R., Gunn, J., Knapp, G., Rockosi, C., \& Blanton, M. 2011, AJ, 142, 116

\bibitem{} Feng, J. Q., \& Gallo, C. F. 2011, Research in Astron. Astrophys., 11, 1429

\bibitem{} Fuchs, B., Jahreiss, H., \& Wielen, R. 1999, Ap\&SS, 265, 175

\bibitem{} Garbari, S., Liu, C., Read, J. I., \& Lake, G. 2012, MNRAS, 425, 1445

\bibitem{} Gilmore, G., Wyse, R. F. G., Norris, J. E. 2002, ApJ, 574, L39

\bibitem{} Golubov, O., Just, A., Bienaym\'e, O., et al. 2013, A\&A, 557, A92

\bibitem{} Honma M., \& Sofue Y. 1996, PASJ, 48, L103

\bibitem{} Kuzmin, G. G. 1952, Tartu Astr. Obs. Publ. 32, 5

\bibitem{} Kuzmin, G. G. 1955, Tartu Astr. Obs. Publ. 33, 3

\bibitem{} Laney, C. D., Joner, M. D., \& Pietrzy\'nski, G. 2012, MNRAS, 419, 1637

\bibitem{} L\'opez-Corredoira, M., \& Betancort-Rijo, J. E. 2004, A\&A, 416, 7

\bibitem{} L\'opez-Corredoira, M., Cabrera-Lavers, A., Garz\'on, F., \& Hammersley, P. L.
2002, A\&A, 394, 883

\bibitem{} L\'opez-Corredoira, M., Cabrera-Lavers, A., Mahoney, T. J., Hammersley, P. L., 
Garz\'on, F., \& Gonz\'alez-Fern\'andez, C. 2007, AJ, 133, 154

\bibitem{} Malkin, Z. M. 2013, Astron. Reports, 57, 128

\bibitem{} Marshall, D. J., Robin, A. C., Reyl\'e, C., Schultheis, M., \& Picaud, S.
2006, A\&A, 453, 635

\bibitem{} McMillan, P. J., \& Binney, J. J. 2010, MNRAS, 402, 934

\bibitem{} Nataf, D. M., Gould, A., Fouqu\'e, P., et al. 2013, ApJ, 769, 88 

\bibitem{} Pietrzy\'nski, G., Gieren, W., \& Udalski, A. 2003, AJ, 125, 2494

\bibitem{} Reid, M. J., Menten, K. M., Brunthaler, A., et al. 2014, arXiv:1401.5377

\bibitem{} Reyl\'e, C., Marshall, D. J., Robin, A. C., \& Schultheis, M. 2009, A\&A, 495, 819

\bibitem{} Roeser, S., Demleitner, M., \& Schilbach, E. 2010, AJ, 139, 2440

\bibitem{} Sch\"onrich, R. 2012, MNRAS, 427, 274

\bibitem{} Sevenster, M. N., Prasenjit, S., Valls-Gabaud, D., \& Fux, R. 1999,
MNRAS, 307, 584

\bibitem{} Sofue, Y., Honma, M., \& Omodaka, T. 2009, PASJ, 61, 227

\bibitem{} Sofue Y. 2011, PASJ, 63, 813

\bibitem{} Wegg, C., \& Gerhard, O. E. 2013, MNRAS, 435, 1874

\bibitem{} Williams, M. E. K., Steinmetz, M., Binney, J., et al. 2013,
MNRAS, 436, 101

\bibitem{} Wu, Z.-Y., Ma, J., \& Zhou, X. 2011, PASP, 12, 1313

\bibitem{} Yaz G\"ok\c ce, E., Bilir, S., \"Ozt\"urkmen, N. D., Duran, \c S., Ak, T., Ak, S., 
\& Karaali, S. 2013, New Astron., 25, 19

\end{thebibliography}
\end{document}